\documentclass[reprint,amsmath,amssymb,aps, prapplied, longbibliography]{revtex4-2}

\usepackage[T1]{fontenc}
\usepackage{graphicx}
\usepackage{subfigure}
\usepackage{multirow}
\usepackage{xcolor,soul}
\usepackage{longtable}
\usepackage{dcolumn}
\usepackage[colorlinks=true,allcolors=blue]{hyperref}
\usepackage{bm}
\usepackage{amsfonts}
\usepackage{amsmath}
\usepackage{amssymb}
\usepackage{svg}
\usepackage{amsthm}
\usepackage{MnSymbol}
\usepackage{enumerate}

\theoremstyle{definition}

\begin{document}

\title{Single-photon-boosted type-I fusion gates}

\author{A.\,A.\,Melkozerov$^{1,2}$}
\email{melkozerov.alex@gmail.com}
\author{S.\,S.\,Straupe$^{3,1,2}$}
\author{M.\,Yu.\,Saygin$^{3,2}$}

\affiliation{\it $^1$ Russian Quantum Center, Bolshoy Boulevard 30, building 1, Moscow, 121205, Russia \\
\it $^2$ Faculty of Physics, M.\,V. Lomonosov Moscow State University, Leninskie Gory 1, Moscow, 119991, Russia \\
\it $^3$ Sber Quantum Technology Center, Kutuzovski prospect 32, Moscow, 121170, Russia}

\date{March, 2026}

\begin{abstract}
Fusion measurements are a key primitive for linear-optical quantum computing and quantum networks. Type-I and type-II fusion gates are widely used to combine small entangled resource states into larger photonic states, but without ancillary resources their success probability is limited to $1/2$. Existing $3/4$-efficient type-I schemes rely on entangled Bell-pair ancillary states, whose preparation is itself probabilistic and resource-intensive. Here we propose a boosted type-I fusion gate using only four ancillary single photons and standard linear-optical primitives. For multiqubit GHZ and graph states, the gate succeeds directly with probability $5/8$, while a distillation protocol converts partially entangled outcomes into additional successful events, raising the total success probability to $11/16$ after one stage and asymptotically to $3/4$. We quantify the practical advantage of this scheme by estimating the photonic resources required for generating representative large entangled photonic states and show that the proposed gate significantly reduces the required overhead compared with existing schemes. These results expand the set of resource-efficient linear-optical primitives and enable a substantial reduction in the resource requirements for scalable photonic quantum computing and quantum communication.
\end{abstract}

\maketitle

\section{Introduction}

Single photons manipulated by linear-optical elements provide one of the leading platforms for photonic quantum technologies, including quantum computing \cite{Knill2001, Kok2007, PsiQuantum2025} and quantum networks \cite{OBrien2009, Wehner2018}. A key limitation of this approach is that in dual-rail encoding entangling operations cannot be implemented deterministically using only passive linear optics.

However, measurement-based and fusion-based approaches provide promising frameworks for universal and fault-tolerant quantum computing in this setting. Scalable architectures \cite{GimenoSegovia2015, Omkar2022, Bartolucci2023, Pankovich2024, Paesani2023} rely on two key primitives: the preparation of small entangled resource states, such as GHZ or graph states, and the implementation of entangling projective measurements, known as \textit{fusions} \cite{Browne2005}. The efficiency of these operations must exceed certain thresholds to enable fault-tolerant error correction.

A successful type-I fusion gate applies the entangling map $|0\rangle\langle 00| + |1\rangle\langle 11|$ to two photonic qubits and measures one of them. Type-II fusion, by contrast, performs a Bell-state measurement on two input qubits and measures both. In the absence of ancillary photonic resources, the success probability of both fusion types is limited to $1/2$ \cite{Calsamiglia2001}.

Improving this success probability is crucial not only at the level of individual gates, but also for the feasibility of scalable linear-optical architectures. In fusion-based quantum computing (FBQC), the nondeterminism of fusion operations can be tolerated provided that their success probability exceeds certain thresholds. For example, these thresholds are $0.57$ and $0.67$ for $(2,2)$-Shor-encoded 4-star and 6-ring fusion networks, respectively \cite{Bartolucci2023}. In percolation-based approaches, representative thresholds are $0.746$ for a two-dimensional 3-GHZ-based lattice and $0.627$ and $0.611$ for three- and four-dimensional lattices, respectively \cite{GimenoSegovia2015,Pant2019}. Thus, even moderate improvements in fusion efficiency can enable new computational schemes or enhance error tolerance \cite{Bartolucci2023, Bombin2023}.

Type-II fusion, which realizes a Bell-state measurement in the dual-rail setting, can be boosted to a $3/4$ success probability using either entangled \cite{Grice2011,Ewert2014,Schmidt2024, Hauser2025} or single-photon \cite{Ewert2014, Bartolucci2021, Schmidt2024, Bayerbach2023, Guo2024} ancillary states. Type-I fusion is particularly attractive for resource-state preparation because it consumes only one of the fused qubits rather than both. However, previously reported $3/4$-efficient type-I schemes rely on ancillary Bell states \cite{Bartolucci2021}. Since such Bell states must themselves be generated probabilistically from at least four single photons \cite{Stanisic2017} in complex linear-optical circuits, they introduce significant resource overhead.

In this work, we show that the success probability of type-I fusion of GHZ states, including Bell states as the two-qubit case, and graph states can be boosted using only single-photon ancillary inputs, which are significantly easier to generate experimentally. Our protocol employs four ancillary single photons and standard linear-optical primitives. It produces direct successful fusion outcomes with probability $5/8$, while a distillation step converts a class of partially entangled outcomes into additional successful events, increasing the total success probability to $3/4$ in the asymptotic limit. To the best of our knowledge, a boosted type-I fusion gate assisted only by single-photon ancillary states has not been reported previously.

We also examine the implications of the proposed gate for resource-state generation. Fusion operations are central to the creation of large entangled photonic states, which serve as key resources for fault-tolerant linear-optical quantum computing, quantum communication, and quantum metrology \cite{Shettell2020,Tth2014,Friis_2017,Farouk2017,Lee2019}. In all-photonic approaches, small entangled states are first generated probabilistically from single photons and subsequently fused into larger states \cite{Zaidi2015,Bartolucci2021,Browne2005,Kieling2007,GimenoSegovia2015, Pankovich2024_flexible, Sahay2023}. Near-deterministic sources can then be realized via multiplexing of many probabilistic generation attempts \cite{Switch_networks2021,Kaneda2026}. 

Here we estimate the average number of single photons required in several multiplexed fusion-based schemes for generating large entangled states and demonstrate that the proposed boosted gate significantly reduces the overall resource cost.

The paper is organized as follows. In Sec.~\ref{sec:preliminaries} we introduce the notation and review standard type-I fusion in dual-rail encoding. In Sec.~\ref{sec:boosted_type_I} we present the boosted gate, classify its detection outcomes, and describe the distillation protocol that raises the total success probability asymptotically to $3/4$. In Sec.~\ref{sec:resource_estimates} we quantify the resulting resource savings for representative state-generation schemes and compare the proposed gate with existing boosted type-I and type-II fusion protocols.
Finally, in Sec.~\ref{sec:conclusion} we discuss the practical significance, limitations of the proposed approach, and prospects for further improvements.

\section{Preliminaries}\label{sec:preliminaries}

In dual-rail encoding, a logical qubit is represented by a single photon occupying one of two optical modes:
\[
|0\rangle = |10\rrangle = \hat{a}_1^{\dagger}|vac\rrangle,
\qquad
|1\rangle = |01\rrangle = \hat{a}_2^{\dagger}|vac\rrangle,
\]
where $\hat{a}_1^{\dagger}$ and $\hat{a}_2^{\dagger}$ are bosonic creation operators and $|vac\rrangle$ is the multimode vacuum state. We use double-ket notation for physical Fock states and ordinary kets for logical qubit states.

\begin{figure}[!h]
    \centering
    \includegraphics[width=0.25\textwidth]{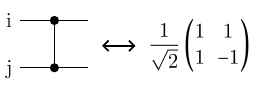}
    \caption{Notation for a balanced $50{:}50$ beam splitter acting on two optical modes.}
    \label{fig:beamsplitter}
\end{figure}

In circuit diagrams, optical modes are represented by horizontal lines. A lossless $m$-mode linear-optical interferometer is described by an $m\times m$ unitary matrix. A balanced beam splitter acting on two modes $i<j$ is represented by a $2\times 2$ Hadamard transformation, and we use the notation shown in Fig.\ref{fig:beamsplitter}.

Let the initial state consist of two initially separated maximally entangled states of the form
\begin{equation}\label{eq:in_state}
    \frac{1}{\sqrt{2}}\Bigl( |A_0\rangle |0\rangle_a + |A_1\rangle |1\rangle_a \Bigr) \otimes \frac{1}{\sqrt{2}} \Bigl(|B_0\rangle |0\rangle_b+ |B_1\rangle |1\rangle_b \Bigr),
\end{equation}
where $\langle A_i|A_j\rangle = \langle B_i|B_j\rangle = \delta_{ij}$. This representation allows one to describe the fusion of the states most relevant to linear-optical quantum computing and communication, including arbitrary graph states, Bell states and multiqubit GHZ states (see Appendix~\ref{sec:AppendixA}). If modes $1$--$2$ encode qubit $a$ and modes $3$--$4$ encode qubit $b$, the corresponding Fock-state form is
\begin{equation}\label{eq:in_state_fock}
    \begin{aligned}
        | \psi^{(in)}\rangle = &\frac{1}{\sqrt{2}}\bigl( |A_0\rangle |1_1 0_2\rrangle + |A_1\rangle |0_1 1_2\rrangle \bigr) \otimes \\
        &\frac{1}{\sqrt{2}} \bigl(|B_0\rangle|1_3 0_4\rrangle + |B_1\rangle|0_3 1_4\rrangle \bigr) \\
        =&\frac{1}{2}\bigl( |A_0 B_0\rangle |1_1 0_2 1_3 0_4\rrangle  + |A_0 B_1\rangle |1_1 0_2 0_3 1_4\rrangle \\
        &+ |A_1 B_0\rangle |0_1 1_2 1_3 0_4\rrangle  + |A_1 B_1\rangle |0_1 1_2 0_3 1_4\rrangle \bigr),
    \end{aligned}
\end{equation}
where tensor products are omitted for compactness.

In the linear-optical setting, measurements are performed with photon-number-resolving detectors (PNRDs). If detectors register a pattern $\vec{n}=(n_1,\ldots,n_m)$ in modes $1,\ldots,m$, the corresponding Kraus operator acting on these modes is
\begin{equation}\label{eq:Kraus}
    K_{\vec{n}} = \bigotimes_{i=1}^{m}|0\rrangle\llangle n_i| = |\vec{0}\rrangle\llangle \vec{n}|.
\end{equation}
If the detectors are preceded by a linear-optical interferometer, the total Kraus operator takes the form
\begin{equation}
    K_{\vec{n}} = |\vec{0}\rrangle\llangle \vec{n}| \; \mathcal{U}(U),
\end{equation}
where $\mathcal{U}(U)$ is the induced Fock-space transformation corresponding to the interferometer matrix $U$. In the present work, each mode contains either zero, one, or two photons. Therefore, it is convenient to introduce an explicit projection onto the corresponding subspace:
\begin{equation}\label{eq:final_Kraus}
    K_{\vec{n}} = |\vec{0}\rrangle\llangle \vec{n}| \; \mathcal{U}(U) \bigotimes_{i=1}^{m} P_i,
\end{equation}
where $P_i = |0_i\rrangle\llangle0_i| + |1_i\rrangle\llangle1_i|$ or $P_i = |0_i\rrangle\llangle0_i| + |2_i\rrangle\llangle2_i|$. Since photodetection destroys all measured photons, we discard the corresponding modes after the measurement.

\subsection{Type-I fusion gates}

\begin{figure}[t]
    \includegraphics[width=0.5\textwidth]{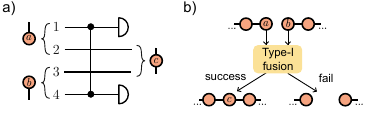}
    \caption{\textbf{Type-I fusion gate}. (a) Linear-optical circuit for the standard type-I fusion gate. (b) A successful event is heralded by the detection of exactly one photon in mode $1$ or mode $4$, implementing the map $|0\rangle\langle 00| \pm |1\rangle\langle 11|$ on the input qubits. Detection of zero or two photons leaves the output states unentangled, corresponding to gate failure.}
    \label{fig:RegularType1}
\end{figure}

Fusion gates were initially introduced as a linear-optical tool for probabilistically connecting separated entangled states \cite{Browne2005}. We briefly review the standard type-I fusion gate acting on qubits $a$ and $b$ of the input state \eqref{eq:in_state}. The corresponding circuit acting on modes $1-4$ is shown in Fig.~\ref{fig:RegularType1}.

The gate outcome is determined by the detection pattern in modes $1$ and $4$. Applying Eq.~\eqref{eq:final_Kraus}, we obtain the Kraus operators
\begin{equation}\label{eq:Type_1_Kraus}
    \begin{aligned}
        &K_{(1,0)} = \frac{1}{\sqrt{2}}\bigl( \llangle 1_1 0_4 | + \llangle 0_1 1_4 | \bigr); \\
        &K_{(0,1)} = \frac{1}{\sqrt{2}}\bigl( \llangle 1_1 0_4 | - \llangle 0_1 1_4 | \bigr); \\
        &K_{(2,0)} = K_{(0,2)} = \frac{1}{\sqrt{2}} \llangle 1_1 1_4 |; \\
        &K_{(0,0)} = \llangle 0_1 0_4 |.
    \end{aligned}
\end{equation}

If exactly one photon is detected in mode $1$ or mode $4$, the gate succeeds and produces the entangled output
\begin{equation}\label{eq:standard_output}
    \begin{aligned}
        |\psi^{(out)}\rangle &= K_{\vec{n}} |\psi^{(in)}\rangle/\sqrt{P_{\vec{n}}} \\
        &= \frac{1}{\sqrt{2}}\bigl( |A_0 B_0\rangle |0_2 1_3\rrangle \pm |A_1 B_1\rangle |1_2 0_3\rrangle \bigr) \\
        &= \frac{1}{\sqrt{2}}\bigl(|A_0 B_0\rangle|0\rangle_c \pm |A_1 B_1\rangle|1\rangle_c\bigr),
    \end{aligned}
\end{equation}
where $|0_2 1_3\rrangle \equiv |0\rangle_c$ and $|1_2 0_3\rrangle \equiv |1\rangle_c$ define the surviving output qubit. The corresponding success probability is
\begin{equation}
    P^{s} = P_{(1,0)} + P_{(0,1)} = \frac{1}{2}.
\end{equation}
If no photons or two photons are detected in modes $1$ and $4$, the gate fails, leaving the output state $|A_0B_1\rangle$ or $|A_1B_0\rangle |1_2 1_3\rrangle$, respectively.

Compared with type-II fusion, type-I fusion can be more resource-efficient because it measures only one of the fused qubits. Its main drawback is that loss detection is less direct, since not all input modes are measured. In many architectures, however, losses induced during type-I fusion can be detected at later stages of the protocol. For example, if type-I gates are used during the preparation of resource states for FBQC protocols\cite{Bartolucci2021}, the resulting states are subsequently fully measured during computation, allowing losses induced by type-I fusion to be treated together with other loss channels \cite{Melkozerov2024, PsiQuantum2025_2}.

\section{Boosting type-I fusion}\label{sec:boosted_type_I}

\begin{figure}[t]
    \includegraphics[width=0.25\textwidth]{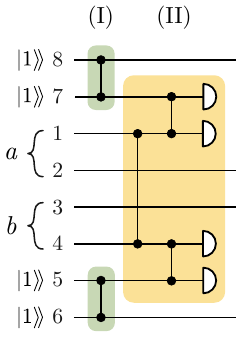}
    \caption{\textbf{Boosted type-I fusion gate using four ancillary single photons}. Part (I) converts the ancillary input $|1111\rrangle$ into two states of the form $(|20\rrangle-|02\rrangle)/\sqrt{2}$. Part (II) interferes the input qubits with these ancillary states and measures modes $1$, $4$, $5$, and $7$. Odd-photon detection events yield the standard type-I fusion output, a subset of four-photon events yields a two-qubit entangled output, and two-photon events can be converted into successful fusion outcomes by the distillation protocol of Fig.~\ref{fig:Distillation}.}
    \label{fig:BoostedType1}
\end{figure}

The standard type-I fusion gate fails because terms associated with $|0_1 0_4\rrangle$ and $|1_1 1_4\rrangle$ in the initial state \eqref{eq:in_state_fock} remain distinguishable at the detection stage. The central idea of boosting is to erase this distinguishability by introducing suitable ancillary states. The scheme shown in Fig.~\ref{fig:BoostedType1} achieves this using four ancillary single photons. This constitutes the key conceptual difference from the boosted type-I protocol of Ref.~\cite{Bartolucci2021}, which relies on ancillary Bell states. 

The possible outputs of the gate are summarized in Table~\ref{table:boosted_fusion_results}. A standard successful fusion outcome occurs directly with total probability $1/2$. In addition, a class of four-photon detection events produces useful entangled two-qubit outputs with probability $1/8$. Finally, when two photons are detected in total, the circuit yields a partially entangled output that can be converted into a standard successful fusion event by a distillation procedure. A single distillation stage contributes an additional success probability of $1/16$, giving a total success probability of $11/16$, whereas the full protocol contributes an additional $1/8$, raising the total success probability to $3/4$.

    \begin{table}[t]
    \renewcommand{\arraystretch}{1.25}
        \begin{tabular}{cccc}
        \hline
                                                      & photons detected             & $P$ & $|\psi^{(out)}\rangle$                \\ \hline
        \multicolumn{1}{c|}{\multirow{3}{*}{\rotatebox[origin=c]{90}{success}}} & 1,3,5  & $1/2$ & $|A_0 B_0\rangle|0\rangle \pm |A_1 B_1\rangle|1\rangle$ \\
        \multicolumn{1}{c|}{} & 4 (some patterns) & $1/8$ & $|A_0 B_1 \rangle|00\rangle \pm |A_1 B_0 \rangle|11\rangle$ \\
        \multicolumn{1}{c|}{} & 2$^\ast$ & $0$--$1/8$ & $|A_0 B_1 \rangle|0\rangle \pm |A_1 B_0 \rangle|1\rangle$  \\ 
        \hline
        \multicolumn{1}{c|}{\multirow{2}{*}{\rotatebox[origin=c]{90}{fail}}} & 2$^\ast$ & $3/16$--$1/16$ & \multirow{2}{*}{$|A_0 B_1\rangle \;\text{or}\; |A_1 B_0\rangle$} \\
        \multicolumn{1}{c|}{} & 0,6,4 (some patterns) & $3/16$ & \\
        \hline
        \end{tabular}
    \caption{Outcome classes of the boosted type-I fusion gate for input states of the form $\bigl( |A_0\rangle |0\rangle_a + |A_1\rangle |1\rangle_a \bigr)\otimes \bigl( |B_0\rangle |0\rangle_b + |B_1\rangle |1\rangle_b \bigr)$, grouped by the total number of photons detected in modes $1$, $4$, $5$, and $7$. For the $n_d=2$ branch, the success probability depends on whether the optional distillation protocol is applied.}
    \label{table:boosted_fusion_results}
    \end{table}
All outcomes are heralded by the detection pattern, and the unmeasured modes are then processed conditionally. The amount of classical feed-forward required depends on the chosen distillation regime. 

Additional outputs arising in the boosted gate are locally Clifford-equivalent to the output of the standard type-I fusion gate, differing only by Pauli $Z$ and $X$ byproduct operators for arbitrary graph states and multiqubit GHZ states (see Appendix~\ref{sec:AppendixA} for details). In measurement-based and fusion-based quantum computing architectures, such byproduct operators do not need to be physically applied. Instead, they are tracked in the classical Pauli frame and absorbed into the bases of subsequent measurements \cite{Omkar2022,Bartolucci2023}. Furthermore, these corrections can be implemented deterministically in dual-rail encoding whenever required. A Pauli $X$ gate is a simple swap of the modes encoding the qubit, while a Pauli $Z$ gate corresponds to a phase shift.

Below we describe the action of the boosted gate in detail. The total input state is $| \psi^{(in)}\rangle \otimes |1_51_61_71_8\rrangle$. Part (I) of the circuit in Fig~\ref{fig:BoostedType1} transforms the ancillary photons according to
    \begin{equation}\label{eq:in_anc}
        \begin{aligned}
            &| \psi^{(in)}\rangle \otimes |\text{anc}\rrangle =  | \psi^{(in)}\rangle \otimes \frac{1}{2} \bigl( |20\rrangle - |02 \rrangle \bigr)\otimes\bigl( |20\rrangle - |02 \rrangle \bigr) \\
            &=\frac{1}{4}\bigl(|A_0B_0\rangle|\mathbf{1}01\mathbf{0}\rrangle + |A_0B_1\rangle|\mathbf{1}00\mathbf{1}\rrangle + |A_1B_0\rangle|\mathbf{0}11\mathbf{0}\rrangle + \\
            &+|A_1B_1\rangle|\mathbf{0}10\mathbf{1}\rrangle \bigr) \otimes\bigl(|\mathbf{2}0\mathbf{2}0\rrangle - |\mathbf{2}0\mathbf{0}2\rrangle- |\mathbf{0}2\mathbf{2}0\rrangle + |\mathbf{0}2\mathbf{0}2\rrangle\bigr).
        \end{aligned}
    \end{equation}

The modes acted upon and measured in part (II) of the circuit are highlighted in boldface. Here and in the following, we omit mode indices for simplicity and list them in increasing order.

For a detection pattern $\vec{n}=(n_1,n_4,n_5,n_7)$, the total number of detected photons $n_d$ is odd for the terms proportional to $|A_0B_0\rangle$ and $|A_1B_1\rangle$, and even for the terms proportional to $|A_0B_1\rangle$ and $|A_1B_0\rangle$.

\subsection{Direct fusion events $(n_d=1,3,5)$}
The total probability of detecting an odd number of photons is $P^{s}_{\text{odd}}=1/2$. The relevant Kraus operators are
    \begin{equation}\label{eq:Type_1_odd_Kraus}
        K_{1 \pm} = \omega_1\bigl( \llangle 1_1 0_4 | \pm \llangle 0_1 1_4 | \bigr)\otimes \llangle\text{ex}_{5,7}|,
    \end{equation}
where $\llangle\text{ex}_{5,7}| = \llangle0_50_7|$; $\llangle0_52_7|$, or $\llangle2_50_7|$; or $\llangle2_52_7|$ for one, three or five photons measured in total respectively.
    \begin{equation}\label{eq:omegas}
        \omega_1 \in \{ 1/2, \sqrt{3}/4,  \sqrt{2}/4, 1/4, \sqrt{6}/8,  \sqrt{3}/8,\sqrt{2}/8\},
    \end{equation}
where a specific value depends on the detection pattern. The output is then
    \begin{equation}\label{eq:boosted_odd}
        \begin{aligned}
            &|\psi^{(out)}_{\vec{n}}\rangle = K_{1 \pm} (| \psi^{(in)}\rangle \otimes |\text{anc}\rrangle)/\sqrt{P_{\vec{n}}}  \\
            &=\frac{1}{\sqrt{2}}\bigl(|A_0B_0\rangle|0_21_3\rrangle \pm |A_1B_1\rangle|1_20_3\rrangle\bigr)\otimes|\text{ex}_{6,8}\rrangle,
        \end{aligned}
    \end{equation}
which is equivalent to the successful output of the standard type-I fusion gate, up to known ancillary by-products.

\subsection{Two-qubit outputs $(n_d=4)$}

If four photons are detected in total, two qualitatively distinct classes of events occur. When all four photons are detected in modes $1$ and $7$, or in modes $4$ and $5$, the gate fails and leaves a separable output. The probability of this event is calculated from Eq.~\eqref{eq:in_anc}, and reads $P_{n_d=4}^f =1/16$. For the complementary class, when two photons are detected in each of the mode pairs $(1,7)$ and $(4,5)$, the circuit produces a useful two-qubit entangled state. The corresponding Kraus operators are
    \begin{equation}\label{eq:boosted_4_success}
            K_{2 \pm} = \omega_2\bigl( \frac{1}{2}(\llangle 1 1 2 0 | \pm \llangle 1 1 0 2 |) \pm \frac{1}{\sqrt{2}}  \llangle 0 0 2 2 |\bigr),
    \end{equation}
where $\omega_2 \in \{ 1/(\sqrt{2}), 1/2,1/(2\sqrt{2}) \}$ depending on the detection pattern, and
    \begin{equation}\label{eq:P_s_4}
            P^{s}_{n_d=4} = 1/8.
    \end{equation}
The output states take the form
    \begin{equation}\label{eq:boosted_4_res}
        \begin{aligned}
            &|\psi^{(out)}_{\vec{n}}\rangle = K_{2 \pm} (| \psi^{(in)}\rangle \otimes |\text{anc}\rrangle)/\sqrt{P_{\vec{n}}}  \\
            &=\frac{1}{2}|A_0B_1\rangle |00 \rrangle \bigl(|02\rrangle \pm |20\rrangle\bigr) \mp \frac{1}{\sqrt{2}}|A_1B_0 \rangle|1100\rrangle.
        \end{aligned}
    \end{equation}
After a balanced beam splitter on modes $6$ and $8$ and, when required, a phase shift on mode $6$, this branch yields the two-qubit entangled output
    \begin{equation}\label{eq:boosted_4_final}
        \begin{aligned}
            |\psi^{(out)}\rangle &= \frac{1}{\sqrt{2}}\bigl(|A_0B_1\rangle |00 11\rrangle \pm |A_1B_0 \rangle|1100\rrangle\bigr),
        \end{aligned}    
    \end{equation}
which encodes a useful two-qubit resource. Output qubits $c_1$ and $c_2$ are encoded in modes $2,6$ and $3,8$, respectively, with $|0\rangle_{c_1}=|1_2 0_6\rrangle$ and $|1\rangle_{c_1}=|0_2 1_6\rrangle$ (and analogously for $c_2$). The output state can therefore be written as
\begin{equation}
|\psi^{(out)}\rangle = \frac{1}{\sqrt{2}}\bigl(|A_0B_1\rangle |0_{c_1}0_{c_2}\rangle
\pm
|A_1B_0\rangle |1_{c_1}1_{c_2}\rangle \bigr).
\end{equation}

By swapping modes $3$ and $6$, qubit $c_1$ becomes encoded in modes $2,3$, matching the encoding of qubit $c$ in the direct fusion branch. In integrated photonic platforms, such a mode swap can be readily implemented using waveguide crossings, which introduce only low optical losses \cite{Johnson2020}. 

By measuring qubit $c_2$ in the Pauli-$X$ basis, the resulting state can then be converted into a standard type-I fusion outcome up to Pauli by-products (see Appendix \ref{sec:AppendixA} for details):

\begin{equation} \label{eq:additional_outputs}
|\psi^{(out)}\rangle = \frac{1}{\sqrt{2}}\bigl(|A_0B_1\rangle |0_{c}\rangle
\pm
|A_1B_0\rangle |1_{c}\rangle \bigr).
\end{equation}

This measurement is implemented using a single balanced beam splitter followed by photon detection in modes $6$ and $8$. The same beam splitter and detectors may also remain in place for the odd-photon direct fusion branches ($n_d=1,3,5$). In these branches, modes $6$ and $8$ contain known ancillary by-product states, which are discarded, and no additional postselection is imposed on the detection in these modes.

\subsection{Partially entangled outcomes and distillation $(n_d=2)$}

When two photons are detected in total, the gate produces partially entangled states. These events occur with probability $3/16$ and are described by Kraus operators
    \begin{equation}\label{eq:boosted_2_Kraus}
        \begin{aligned}
            &K_{3 \pm} = \omega_3\bigl( \frac{\sqrt{2}}{\sqrt{3}}\llangle 0 0 0 2| \pm \frac{1}{\sqrt{3}} \llangle 1 1 0 0 | \bigr), \\ 
            &K_{4 \pm} = \omega_4\bigl( \frac{\sqrt{2}}{\sqrt{3}}\llangle 0 0 2 0| \pm \frac{1}{\sqrt{3}} \llangle 1 1 0 0 | \bigr),
        \end{aligned}
    \end{equation}
where $\omega_{3,4} \in\{ \sqrt{3}/2, \sqrt{3}/(2\sqrt{2}) \}$ depending on the detection pattern. This leads to output states
    \begin{equation}\label{eq:boosted_2}
        \begin{aligned}
            |\psi^{(out)}_{\vec{n}}\rangle = &K_{3,4 \pm} (| \psi^{(in)}\rangle \otimes |\text{anc}\rrangle)/\sqrt{P_{\vec{n}}}  \\
            =&\frac{1}{\sqrt{3}}|A_0B_1\rangle |0 0 2 2 \rrangle  \pm \frac{\sqrt{2}}{\sqrt{3}}|A_1B_0 \rangle|1 1 2 0\rrangle, \; \text{or} \\
            =&\frac{1}{\sqrt{3}}|A_0B_1\rangle |0 0 2 2  \rrangle  \pm \frac{\sqrt{2}}{\sqrt{3}}|A_1B_0 \rangle|1 1 0 2\rrangle.
        \end{aligned}    
    \end{equation}

\begin{figure}[h]
        \includegraphics[width=0.47\textwidth]{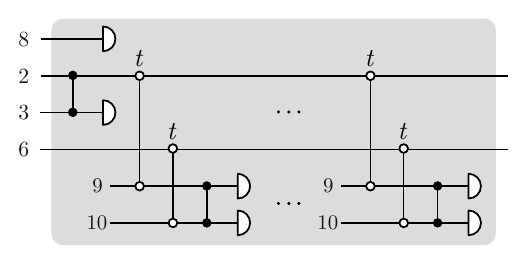}
        \caption{\textbf{Distillation protocol for the $n_d=2$ outputs of the boosted type-I fusion gate}. The circuit consists of balanced beam splitters and beam splitters with equal transmission coefficient $t$. A successful distillation round is heralded by vacuum detection in mode $3$ and single-photon detection in modes $9$ and $10$. If no photons are detected in modes $3$, $9$, and $10$, a part of the procedure can be repeated, until one or two photons are detected in modes $9$ and $10$.}
        \label{fig:Distillation}
    \end{figure}

These states can be distilled into standard successful fusion outputs. For the second outcome, the corresponding circuit is shown in Fig.~\ref{fig:Distillation}; the other outcome is treated analogously after swapping modes $6$ and $8$.

A single distillation stage succeeds when no photons are detected in mode $3$ and exactly one photon is detected in mode $9$ or $10$, yielding
    \begin{equation}\label{eq:distilled}
        \begin{aligned}
            |\psi^{(out)}\rangle &= \frac{1}{\sqrt{2}} \bigl(|A_0 B_1 \rangle|0_2 1_{6}\rrangle \pm |A_1 B_0 \rangle|1_2 0_{6}\rrangle \bigr),
        \end{aligned}    
    \end{equation}
which corresponds to the successful output of the gate \eqref{eq:additional_outputs}. 
If two photons are detected in mode $3$, or in modes $9$ or $10$, the gate fails.

If no photons are detected in modes $3$, $9$, and $10$, the procedure can be repeated until one or two photons are detected in modes $9$ and $10$ (see Fig.~\ref{fig:Distillation}). If exactly one photon is detected, the procedure is stopped, yielding the successful output state~\eqref{eq:distilled}.

The repeated version of the protocol is closely related to the bleeding procedure of Ref.~\cite{Bartolucci2021}, while the single-stage version is similar to the fusion protocol used for primate states~\cite{Bartolucci2021,Fldzhyan2025}.

The contribution of this branch to the total fusion success probability is
    \begin{equation}\label{eq:p_succ_2}
            P^s_{n_d=2} = P_{n_d = 2} \cdot P_{dist} = \frac{3}{16} \cdot \frac{4t(1-t^{2k})}{3(1+t)},
    \end{equation}
where $k$ is the maximum number of distillation stages. In the limit $k\rightarrow\infty$ and $t\rightarrow 1$,
    \begin{equation}\label{eq:p_succ_2_max}
            P^s_{n_d=2} = 1/8,
    \end{equation}
so that the total gate success probability reaches $P^{s}=3/4$. If only one distillation stage is used with balanced beam splitters, $t=1/2$, then
    \begin{equation}\label{eq:p_succ_2_one}
            P^s_{n_d=2} = 1/16,
    \end{equation}
yielding a total success probability of $P^{s}=11/16$.

One may alternatively replace the repeated part of the protocol after heralding vacuum in mode $3$ with a single balanced beam-splitter on modes $2$ and $6$, and measure these modes. With this modification, the ($n_d = 2$) branch will immediately yield type-II-like outputs of the form 
    \begin{equation}
        \begin{aligned}
            |\psi^{(out)}\rangle &= \frac{1}{\sqrt{2}} \bigl(|A_0 B_1 \rangle\pm |A_1 B_0 \rangle\bigr)
        \end{aligned}    
    \end{equation}
with a success probability of $P^{s, II}_{n_d=2} = 1/8$.

\subsection{Failure events$(n_d=0,6)$}

If zero or six photons are detected in total, the gate fails. These branches occur with probability $1/16$ each and leave separable outputs  $|\psi^{(out)} \rangle = |A_1B_0\rangle|1122\rrangle $ and $|\psi^{(out)}\rangle = |A_0B_1\rangle|0000\rrangle $ respectively. 

Together with the unsuccessful subsets of the $n_d=2$ and $n_d=4$ branches, they account for the full complement of failure events of the boosted protocol.

\section{Resource estimates for fusion-based state generation}\label{sec:resource_estimates}

\begin{table*}[ht]
\renewcommand{\arraystretch}{1.25}
\setlength{\tabcolsep}{4pt}
\begin{tabular}{c|ccc|c} 
\hline
& \multicolumn{3}{c|}{$P_{3GHZ} = 1/32$} & \multicolumn{1}{l}{$P_{3GHZ} = 1/4$} \\ \cline{2-5}

Fusion gate                & \multicolumn{1}{l|}{$P_{Bell}=1/8$} & \multicolumn{1}{l|}{$P_{Bell}=3/16$} & $P_{Bell}=1/4$                      & $P_{Bell}=2/3$                      \\ \hline
Standard type-II ($P^s=1/2$) \cite{Browne2005} 
& \multicolumn{1}{c|}{768} & \multicolumn{1}{c|}{768}  & \multicolumn{1}{c|}{768} & 96 \\

Single-photon-boosted type-II ($P^s=5/8$) \cite{Ewert2014} 
& \multicolumn{1}{c|}{618} & \multicolumn{1}{c|}{618}  & 618  & 80  \\

Bell-state-boosted type-II ($P^s=3/4$) \cite{Grice2011}
& \multicolumn{1}{c|}{555} & \multicolumn{1}{c|}{540}  & 533  & 72  \\

Single-photon-boosted type-II ($P^s=3/4$) \cite{Ewert2014}
& \multicolumn{1}{c|}{517} & \multicolumn{1}{c|}{517}  & 517  & 69  \\

Standard type-I ($P^s=1/2$) \cite{Browne2005}
& \multicolumn{1}{c|}{320 \cite{Bartolucci2021}} & \multicolumn{1}{c|}{213}  & 160  & 60  \\

Bell-state-boosted type-I ($P^s=3/4$) \cite{Bartolucci2021}
& \multicolumn{1}{c|}{205} & \multicolumn{1}{c|}{137}  & 102  & 38  \\

Single-photon-boosted type-I ($P^s=5/8$)
& \multicolumn{1}{c|}{\textbf{197}} & \multicolumn{1}{c|}{\textbf{136}}  & 105  & 48  \\

Single-photon-boosted type-I ($P^s=11/16$)
& \multicolumn{1}{c|}{--} & \multicolumn{1}{c|}{--}  & \textbf{92}  & 42  \\

Single-photon-boosted type-I ($P^s=3/4$)
& \multicolumn{1}{c|}{--} & \multicolumn{1}{c|}{--}  & --  & \textbf{37} \\

\hline
\end{tabular}
\caption{Average number of single photons required for multiplexed near-deterministic generation of four-qubit GHZ state with various fusion gates for different efficiencies of seed state generators. Target state is generated either by fusing two 3-qubit GHZ states with a single type-II fusion gate, or by fusing three Bell states with two type-I gates. Seed and ancillary Bell states are generated in heralded schemes from four single photons, 3-GHZ states -- from six single photons. We suppose one-stage and full distillation is available for the proposed single-photon-boosted type-I gate when corresponding distillation and bleeding procedures are enabled for seed state generators. The best performing gates for every combination of seed states generators are highlighted in bold.}
\label{table:comparison}
\end{table*}

In this section, we estimate the average number of single photons required for near-deterministic generation of representative photonic resource states in multiplexed schemes. We compare the proposed gate to other fusion gates in Table~\ref{table:comparison} and demonstrate its advantage for resource-state generation. 
Next, we quantify the savings enabled by the proposed gate in different schemes for resource-state generation in Table~\ref{table:resources}. As a resource metric, we use the average number of single-photon inputs required to obtain one successful target state.

We call a resource state generation scheme the sequence of operations that transforms single-photon states into target entangled states. Resource states are commonly generated by first preparing small entangled seed states and then fusing them into larger states \cite{Zaidi2015,Bartolucci2021,Browne2005,Kieling2007,GimenoSegovia2015, Pankovich2024_flexible, Sahay2023}. 

Several proposals exist for the heralded generation of dual-rail Bell states \cite{Zhang2008, Forbes2025,Carolan2015,Bartolucci2021,Fldzhyan2021,gimeno2016} and 3-qubit GHZ states \cite{Forbes2025,Varnava2008,Gubarev2020,Bhatti2025,gimeno2016} from single photons. Among other parameters, they differ by the number of initial single photons and the target state generation success probabilities.


Scalable architectures typically require non-probabilistic operations at least at some stages of the protocol. Fusion-based quantum computing, for example, assumes near-deterministic creation of resource states. Multiplexing can be used to overcome the probabilistic nature of entangling operations \cite{Switch_networks2021,Kaneda2026}. It assumes that a sufficiently large number of probabilistic operations are executed in parallel, and only successful outcomes are selected and routed forward.

To compare different state-generation strategies, we adopt the notion of perfectly resource-efficient multiplexing \cite{Bartolucci2021}. At each stage of a sequential protocol, a large reservoir of $N$ identical probabilistic operations is executed in parallel, and all $M$ successful outputs are routed to the next stage. If a given operation requires $n_0$ single photons per attempt and succeeds with probability $p_0$, then in the limit $N \rightarrow \infty$, $M = Np_0$, and the average number of single photons required per successful output is $(Nn_0)/(Np_0) = n_0/p_0$. This model provides a convenient architecture-level estimate of the resource overhead.

\begin{figure*}[ht]
        \includegraphics[width=1.0\textwidth]{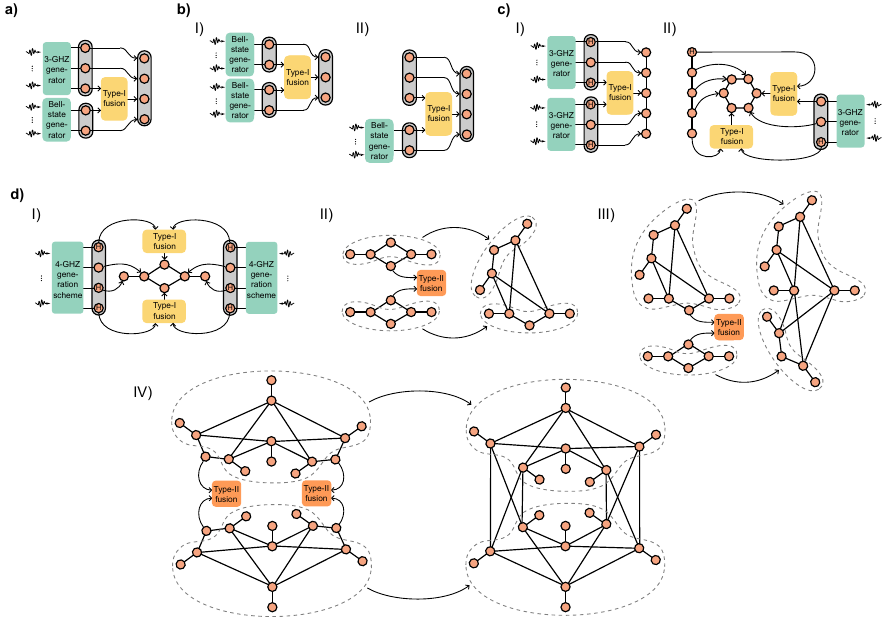}
        \caption{\textbf{Representative fusion-based schemes for creation of photonic resource states}. (a) A four-qubit GHZ state generated from one Bell state and one three-qubit GHZ state using a single type-I fusion gate.  (b) A four-qubit GHZ state generated from three Bell states using two type-I fusion gates. (c) A six-qubit ring graph state, up to local single-qubit operations, generated from three three-qubit GHZ states followed by single-qubit Hadamard gates on the marked qubits and three type-I fusion gates. (d) A twenty-four-qubit (2,2) Shor-encoded six-ring graph state generated from twelve four-qubit states using twelve type-I and six type-II fusion gates.}
        \label{fig:Resource_state_generation}
    \end{figure*}

Fig.~\ref{fig:Resource_state_generation} depicts modular schemes for generating various entangled states based on type-I fusion gates. We assume multiplexing at both the initial state-generation stage and at each subsequent fusion stage, as shown in the figure. 

Table~\ref{table:comparison} summarizes the comparison of the performance of the proposed boosted gate with the standard type-I gate, Bell-state-boosted type-I gate \cite{Bartolucci2021} and three versions of boosted type-II gates in the schemes for creation of representative four-qubit GHZ states. 

Then we estimate the average photon costs for every entangled-state generation scheme from Fig.~\ref{fig:Resource_state_generation} using the unboosted type-I fusion gate and the three variants of the proposed boosted gate. Table~\ref{table:resources} lists the corresponding results.

\subsection{Seed state generation}

Bell states can be generated from minimum four single photons \cite{Stanisic2017}. An eight-mode scheme \cite{Zhang2008,  Stanisic2017, Bartolucci2021} can generate one of the four Bell states with a total $P_{Bell} =1/8$ success probability. When additional $(|1100\rrangle \pm |0011\rrangle)/\sqrt{2}$ states accepted as successful outcomes, it reaches a total $P_{Bell} =3/16$ success probability, highest among other four-photon Bell-state generation schemes \cite{Forbes2025}.

Three-qubit GHZ states can be generated from six single photons with a total success probability of $P_{3GHZ} = 1/32$ \cite{Varnava2008}, the highest among known six-photon schemes \cite{Forbes2025}.

Some unsuccessful outputs of $3/16$-efficient Bell state generation scheme are non-maximally entangled two-qubit states. These states can be converted into successful Bell-state outputs, using a distillation procedure, reaching a total success probability $P_{Bell} =1/4$ \cite{Bartolucci2021}. 

This operation is no less complex in terms of active elements than one-stage distillation, proposed for the boosted type-I gates presented here. 
Thus, we suppose that one-stage distillation is available and the proposed gate can operate at $P^s = 11/16$ success probability when $P_{Bell} = 1/4$ is achievable. 

The success probability of seed state generators can be further enhanced by the bleeding procedure \cite{Bartolucci2021}. It assumes the repeated “weak” photon subtraction until two photons are detected. We use a similar procedure for what we call full distillation in the limit $k \rightarrow \infty$, that allows to reach the total success probability of the fusion gate $P^{s} = 3/4$. 

The combination of distillation and bleeding allows to reach $P_{Bell} = 2/3$ success probability for the Bell state generator \cite{Bartolucci2021}. The success probability of the 3-GHZ generation scheme can also be increased using bleeding, achieving $P_{3GHZ} = 1/4$ \cite{Bartolucci2021}. Since this procedure is no less complex than the proposed method to achieve a total success probability of the type-I fusion gate $P^{s} = 3/4$, we suppose it available when $P_{Bell} = 2/3$ and $P_{3GHZ} = 1/4$ is achievable. 
    \begin{table*}[ht]
    \renewcommand{\arraystretch}{1.25}
    \setlength{\tabcolsep}{4pt}
    \begin{tabular}{c|c|c|c|c}
    \hline
    Scheme
    & \begin{tabular}[c]{@{}c@{}}Standard gate \\ $P^{s}=1/2$\end{tabular}
    & \begin{tabular}[c]{@{}c@{}}Direct boost \\ $P^{s}=5/8$\end{tabular}
    & \begin{tabular}[c]{@{}c@{}}One-stage distillation\\ $P^s=11/16$\end{tabular}
    & \begin{tabular}[c]{@{}c@{}}Full distillation\\$P^{s}=3/4$\end{tabular}\\
    \hline
    (a) $|1\rrangle^{\otimes10} \rightarrow |\text{GHZ}_2\rangle|\text{GHZ}_3\rangle\rightarrow|\text{GHZ}_4\rangle$ \cite{Bartolucci2021}
    & $427$ 
    & $348$
    & $308$
    & $45$ \\
    (b) $|1\rrangle^{\otimes12} \rightarrow |\text{GHZ}_2\rangle^{\otimes 3} \rightarrow|\text{GHZ}_4\rangle$ \cite{Bartolucci2021}
    & $213$
    & $136 \: / \: 160$
    & $92 \: / \: 105$
    & $37 \: / \: 42$ \\
    (c) $|1\rrangle^{\otimes18} \rightarrow |\text{GHZ}_3\rangle^{\otimes 3} \rightarrow|\text{G}_6\rangle$ \cite{Sahay2023}
    & $3\:840$
    & $2\:097$
    & $1\:615$
    & $178$ \\
    (c) $|1\rrangle^{\otimes24} \rightarrow |\text{GHZ}_2\rangle^{\otimes 6}\rightarrow |\text{GHZ}_3\rangle^{\otimes 3}  \rightarrow|\text{G}_6\rangle$
    & $1\:708$
    & $836$
    & $460$
    & $161$ \\
    (d) $|1\rrangle^{\otimes96} \rightarrow |\pi_2\rangle^{\otimes 24} \rightarrow |\text{GHZ}_4\rangle^{\otimes12}\rightarrow|\text{G}_{24}\rangle$ \cite{Wein2025}
    & $197\:760  \: / \: 237\:312$
    & $27\:915\: / \: 30\:454 $
    & $23\:081 \: / \: 25\:180 $
    & $19\:404 \: / \: 21\:168$  \cite{Wein2025} \\
    (d) $|1\rrangle^{\otimes144} \rightarrow |\text{GHZ}_2\rangle^{\otimes 36} \rightarrow |\text{GHZ}_4\rangle^{\otimes12}\rightarrow|\text{G}_{24}\rangle$
    & $136\:533 \: / \: 168\:840 $
    & $12\:484 \: / \: 13\:620  $
    & $7\:144 \: / \: 7\:795 $
    & $2\:622 \: / \: 2\:861 $ \\
    \hline
    \end{tabular}
    \caption{Average number of input single photons required for multiplexed near-deterministic generation of 4-qubit GHZ, 6-qubit ring, and 24-qubit $(2,2)$ Shor-encoded 6-ring graph states using standard and boosted type-I fusion gates. Results are shown for boosted gates and seed state generators with no additional distillation ($k=0, P_{Bell} = 3/16, P_{3GHZ} = 1/32$), one distillation stage ($k=1, P_{Bell} = 1/4, P_{3GHZ} = 1/32$), and full distillation and bleeding ($k=\infty, P_{Bell} = 2/3, P_{3GHZ} = 1/4$). For strategy (b) for 4-qubit GHZ generation, we first present results assuming that two-qubit outcomes of the boosted gate are utilized. For the $(2,2)$ Shor-encoded 6-ring state, we show estimates for the scheme shown in Fig.~\ref{fig:Resource_state_generation} and for the original proposal in Ref.~\cite{Wein2025}.}
    \label{table:resources}
    \end{table*}
\subsection{Comparison of different fusion gates}\label{sec:comparison}

Here we compare the performance of different fusion gates for creation of four-qubit GHZ states. These states are relevant, for example, as resource states for the 4-star network in fusion-based quantum computing \cite{Bartolucci2023}.

It can be generated from two 3-GHZ states, fusing them with a single type-II gate. Two main strategies for boosting type-II gates to $P^s=3/4$ success probability rely on ancillary Bell states \cite{Grice2011} or four single photons \cite{Ewert2014}. The scheme of Ref.~\cite{Ewert2014} can also work in a two-single-photon regime, yielding a success probability of $P^s = 5/8$.

Using type-I gates, three Bell states can be first created and then fused together, forming a four-qubit GHZ state. This scheme, shown in Fig.~\ref{fig:Resource_state_generation}(b), demonstrates the highest efficiency among type-I strategies, and also makes use of the two-photon outcomes of our gate and the Bell-state-boosted type-I gate, proposed in Ref.~\cite{Bartolucci2021}, see Sec.~\ref{sec:4-ghz}.

Ref.~\cite{Bartolucci2021} also proposed a strategy for 4-qubit GHZ state generation based on so-called primate states, with an average cost of approximately $309$ single photons with standard fusion gates. Although this strategy might also benefit from the application of techniques such as boosting or bleeding, its consideration is beyond the scope of the present work.

Estimations of the performance of different fusion gates for multiplexed near-deterministic creation of four-qubit GHZ state are presented in Table~\ref{table:comparison}. We suppose that ancillary states for Bell-state-boosted gates \cite{Grice2011, Bartolucci2021} are created by the same multiplexed Bell state generators as for the seed states. 

In the entangled-state generation scenarios considered here, the proposed type-I gate achieves the lowest average single-photon cost among the studied fusion gates. The Bell-state-boosted type-I gate \cite{Bartolucci2021} demonstrates comparable results, but is still outperformed by the proposed gate even when the full bleeding procedure is allowed for ancillary Bell-state preparation. Moreover, it requires an additional multiplexing stage to prepare the ancillary Bell states.
This advantage becomes even more pronounced when two-photon outputs for both boosted type-I gates are not used, which is the case for other entangled state generation schemes in Fig.~\ref{fig:Resource_state_generation}.

Next we demonstrate how the proposed gate may contribute to the resource savings for the creation of different resource states for quantum computing and quantum communication.

\subsection{4-qubit GHZ states}\label{sec:4-ghz}

Four-qubit GHZ states can be generated directly from eight single photons without intermediate multiplexing \cite{gimeno2016}. The corresponding success probability is $1/128$, which implies an average cost of $1024$ single photons per successful 4-qubit GHZ state.

We consider two generation strategies based on type-I fusion gates. The first, shown in Fig.~\ref{fig:Resource_state_generation}(a), creates one Bell state and one 3-qubit GHZ state from four and six single photons respectively, and fuses them using a single type-I gate. The second, shown in Fig.~\ref{fig:Resource_state_generation}(b) creates three Bell states from single photons and combines them using two type-I fusion gates. In this scheme when boosted gates are applied, we can make use of the outcomes, that save two photons in the output (see Table~\ref{table:boosted_fusion_results}). In this case only one type-I fusion gate at the stage (b1) is enough when it measures exactly four photons in total. 

The average photon costs for the multiplexed strategies are listed in Table~\ref{table:resources} for unboosted type-I fusion and for the three variants of the proposed boosted gate, together with the adoption of two-qubit cases for the second scheme. In both constructions, increasing the success probability of the fusion primitive produces a substantial reduction in the total overhead.

\subsection{6-ring graph states}

Six-qubit ring graph states are the resource states of the 6-ring fusion network, which offers improved robustness in FBQC \cite{Bartolucci2023}. We consider two ways to generate such states up to local single-qubit gates.

The first strategy, shown in Fig.~\ref{fig:Resource_state_generation}(c), begins with three 3-qubit GHZ states generated from six single photons each, followed by local Hadamard gates on selected qubits and three type-I fusion operations \cite{Sahay2023}. The second strategy begins with six Bell states, fuses them pairwise into three 3-qubit GHZ states, and then applies the same fusion step as in the first strategy.

The numerical values in Table~\ref{table:resources} show that the benefit of improved fusion success probability becomes even more pronounced for larger target states. In particular, even the transition from unboosted type-I fusion to the directly boosted gate without additional distillation reduces the average single-photon cost by roughly a factor of two for the six-ring constructions considered here. This illustrates the main architectural motivation for boosted fusion: even moderate improvements at the level of a single fusion gate, achieved with easily prepared ancillary states, can compound into substantial resource savings in large-scale resource-state generation schemes.

\subsection{$(2,2)$-Shor-encoded 6-ring resource states}

The $(2,2)$ Shor encoding substantially improves the loss tolerance of the 6-ring fusion network in the FBQC protocol \cite{Bartolucci2021}, bringing the corresponding loss thresholds close to values that are nearly accessible with current integrated photonic technology \cite{Melkozerov2024}.

Figure~\ref{fig:Resource_state_generation}(d) shows an all-photonic, fusion-based scheme for generating this 24-qubit resource state, following Ref.~\cite{Wein2025}. In this construction, four-qubit resource states are first prepared and then fused using a sequence of type-I and type-II fusion gates. In the original proposal \cite{Wein2025}, stages (d2) and (d3) are assumed to be performed simultaneously. However, if each stage is multiplexed independently, a higher overall efficiency can be achieved. Table~\ref{table:resources} therefore presents resource estimates for both variants of the scheme.

In Ref.~\cite{Wein2025}, the four-qubit graph states, which are local-unitary equivalent to four-qubit GHZ states, are assumed to be generated with an average cost of approximately 309 single photons by using the primate-state scheme proposed in Ref.~\cite{Bartolucci2021}. In Table~\ref{table:resources}, we also include estimates for the four-qubit GHZ-state generation scheme shown in Fig.~\ref{fig:Resource_state_generation}(b), using the same type-I fusion gates as those employed in stage (d1).

We note that Ref.~\cite{Wein2025} effectively assumes a $3/4$ success probability for type-I fusion by analogy with the four-single-photon boosting scheme for type-II fusion introduced in Ref.~\cite{Ewert2014}. However, Ref.~\cite{Ewert2014} does not propose a $3/4$-efficient type-I fusion scheme. In the present work, we explicitly demonstrate that type-I fusion can be boosted from $1/2$ up to $3/4$ using four ancillary single photons. Accordingly, in our estimates, all boosted type-II fusion gates are assumed to achieve $3/4$ success probability via the scheme of Ref.~\cite{Ewert2014}, while type-I fusion gates are assigned the success probabilities listed in Table~\ref{table:resources}.

Our estimations show that even in the mixed strategies with both type-I and type-II gates, the proposed boosting technique provide significant resource savings.

\section{Discussion and conclusion}\label{sec:conclusion}
Linear-optical quantum algorithms rely on the effective realization of two key components: the creation of entangled resource states and entangling measurements, both of which are inherently probabilistic in linear optics. Fusion gates represent a central ingredient in the most advanced photonic quantum algorithms. Here, we have presented a new approach to boosting their success probability, resulting in a significant improvement in multiplexed resource-state generation and, consequently, a substantial reduction in the resource requirements for universal quantum algorithms in linear-optical platforms. 

The proposed boosted type-I fusion gate achieves a total success probability of $75\%$ for Bell states, multiqubit GHZ states, and graph states using only four ancillary single photons and standard linear-optical primitives. The gate succeeds directly with probability $62.5\%$, while a distillation procedure increases this to $68.75\%$ after one stage and to $75\%$ in the asymptotic limit. To our knowledge, this is the first type-I fusion protocol to surpass the $50\%$ limit using only single photons, without requiring entangled ancillary states.

Since the introduction of boosted linear-optical Bell-state measurements (type-II fusion gates) surpassing the $1/2$ success-probability limit using entangled and single-photon ancillary states \cite{Ewert2014, Grice2011}, a wide range of photonic quantum algorithms has been proposed \cite{Bartolucci2023, GimenoSegovia2015, Pant2019}. In contrast, for type-I fusion gates, success-probability enhancement has so far only been achieved using ancillary Bell states \cite{Bartolucci2021}, whose preparation is itself a highly nontrivial probabilistic task. In this work, we close this gap by introducing a single-photon-based boosting scheme for type-I fusion and demonstrate 
its resource efficiency in the entangled-state-generation schemes considered here.

Similar ancillary states were previously proposed for boosting type-II fusion gates \cite{Ewert2014}. Those gates, however, employ them in a fundamentally different manner. In the scheme of Ref.~\cite{Ewert2014}, all input modes are measured, realizing a Bell-state measurement with a success probability of $5/8$ using a single ancillary state of the form $(|20\rangle \pm |02\rangle)/\sqrt{2}$. Using two such ancillary states further increases the success probability to $3/4$. In contrast, the gate proposed here produces an additional successful two-photon branch with probability $1/8$ only when all four ancillary photons are available. If the ancillary resources are restricted to a single state $(|20\rangle - |02\rangle)/\sqrt{2}$, the additional outcomes correspond to partially entangled states and must therefore undergo a distillation step before contributing to the overall success probability.

We have shown that, across several fusion-based resource-generation protocols, the proposed gate substantially reduces the average number of single photons required to generate large entangled states. These savings become more pronounced with increasing system size, as the improved fusion success probability compounds over multiple stages.

According to our calculations, approximately $2622$ single photons are required on average to nearly deterministically generate an entangled 24-qubit (2,2)-Shor encoded 6-ring state in an ideal lossless all-photonic scheme using the proposed gates, which is far below the 
$21\:170$ photon estimate reported in Ref.~\cite{Wein2025}. The corresponding strategy, however, requires more extensive active routing at the fusion stages, which may be experimentally costly.

Our analysis focuses on all-photonic architectures based on single-photon inputs. Experimentally, both heralded generation of small entangled states \cite{Carolan2015, Cao2024, Skryabin2025, Chen2024, Maring2024} and boosted fusion measurements \cite{Bayerbach2023, Guo2024, Hauser2025} have already been demonstrated. Rapid progress is also being made in approaches that combine fusion operations with entangled states generated by quantum emitters \cite{Chan2025, Wein2025}, where the proposed gates may also prove beneficial.

Throughout this work, we have assumed ideal, lossless circuits. In practice, additional multiplexing stages and more complex gates introduce extra optical loss, which may affect the total efficiency of the schemes. An assessment of these trade-offs should be taken into account for experimental implementations.

Natural extensions of this work include the use of larger ancillary states to achieve higher success probabilities, as well as generalizations to multi-qubit fusion gates \cite{Pankovich2024_flexible}.

Finally, the proposed gates may also enable more efficient generation of entangled states beyond the stabilizer formalism. While most measurement-based protocols rely on graph or cluster states, broader classes of resource states, such as weighted graph states and GHZ-like states \cite{Gross2007, Webster2022, Zakaryan2025, Yamazaki2025}, have recently been proposed as promising resources for quantum algorithms. These states can also be generated using fusion-based approaches \cite{Fldzhyan2025, Melkozerov2026}, and improved fusion efficiencies may enhance the corresponding protocols.

\section{Acknowledgments}
The authors’ work on the results presented in Sec. III was
supported by Rosatom in the framework of the Roadmap for Quantum computing (Contract № 868/1759-D dated 3 October 2025 and Contract №11-2025/1 dated 14 November 2025). 

A. Melkozerov is a recipient of a scholarship from the Theoretical Physics and Mathematics Advancement Foundation (BASIS) (No. 24-2-2-5-1) (results presented in Sec. IV)

\appendix

\section{Additional outputs}\label{sec:AppendixA}

The representation of initial states in Eq.~\eqref{eq:in_state} allows one to describe many entangled resource states relevant for quantum computing and quantum communication protocols, including all states studied in Sec.~\ref{sec:resource_estimates}. In this appendix we show that, for these states, the additional successful outputs of the boosted fusion gate are locally Clifford-equivalent to the standard type-I fusion output.

Let $|B\rangle$ be a Bell state or a multiqubit GHZ state. Then $
|B_0\rangle = |0\ldots0\rangle, |B_1\rangle = |1\ldots1\rangle$,
and analogously for the second state $|A\rangle$. The logical flip $|B_0\rangle \leftrightarrow |B_1\rangle$ is therefore implemented by a product of Pauli-$X$ gates on all qubits of the state $|B_i\rangle$, since $|1\rangle = X|0\rangle$. Thus, the output of the additional fusion branch \eqref{eq:additional_outputs} is equivalent to the standard fusion output \eqref{eq:standard_output} up to a product of Pauli-$X$ gates on the surviving qubits of system $B$:
\begin{equation}
\begin{aligned}
    &|\psi^{(out)}_1\rangle =\prod_{i \in B} X_i \; |\psi^{(out)}_2\rangle  \\
    &=\prod_{i \in B} X_i \; \frac{1}{\sqrt{2}} (|A_0 B_1\rangle |0\rangle_c \pm |A_1 B_0\rangle |1\rangle_c) \\
    & = \frac{1}{\sqrt{2}} (|A_0 B_0\rangle |0\rangle_c \pm |A_1 B_1\rangle |1\rangle_c).
\end{aligned}
\end{equation}

Let now $|B\rangle$ be an arbitrary graph state associated with a graph $(V,E)$:
\begin{equation}
    |B\rangle = \prod_{(i,j) \in E} CZ_{(i,j)} \; |+\rangle^{\otimes n},
\end{equation}
where $|+\rangle = H|0\rangle = (|0\rangle + |1\rangle)/\sqrt{2}$. If we single out qubit $b$, this state can be written as
\begin{equation}
\begin{aligned}
    |B\rangle &= \prod_{i:(i,b) \in E} CZ_{i,b} \;  (|B_0\rangle|+\rangle_b) \\
    &= \prod_{i:(i,b) \in E} CZ_{i,b} \;  \frac{1}{\sqrt{2}}\bigl(|B_0\rangle|0\rangle_b + |B_0\rangle|1\rangle_b\bigr)\\
    &=\frac{1}{\sqrt{2}}\Bigl(|B_0\rangle|0\rangle_b + \bigl[ \prod_{i:(i,b) \in E} Z_{i}|B_0\rangle \bigr] \;|1\rangle_b \Bigr),
\end{aligned}
\end{equation}
where
\begin{equation}
|B_0\rangle =
\prod_{\substack{(i,j)\in E \\ i,j\neq b}}
CZ_{(i,j)} \; |+\rangle^{\otimes(n-1)}
\end{equation}
is the graph state of all qubits except the singled-out qubit $b$, with all edges incident on $b$ removed. Therefore the graph state takes the form \eqref{eq:in_state}, with
\begin{equation}
|B_1\rangle =
\prod_{i:(i,b) \in E} Z_i |B_0\rangle .
\end{equation}
An analogous decomposition can be written for the state $|A\rangle$. It immediately follows that the two successful fusion outputs are related by a product of local Pauli-$Z$ corrections on the neighbors of the fused qubit $b$:
\begin{equation}
\begin{aligned}
    &|\psi^{(out)}_1\rangle =\prod_{i:(i,b) \in E} Z_{i} \; |\psi^{(out)}_2\rangle  \\
    &=\prod_{i:(i,b) \in E} Z_{i} \; \frac{1}{\sqrt{2}}(|A_0 B_1\rangle |0\rangle_c \pm |A_1 B_0\rangle |1\rangle_c) \\
    & = \frac{1}{\sqrt{2}}(|A_0 B_0\rangle |0\rangle_c \pm |A_1 B_1\rangle |1\rangle_c).
\end{aligned}
\end{equation}

The transition between different successful outputs of the fusion gate is therefore a local Clifford correction. In measurement-based and fusion-based quantum computing architectures, such byproducts are tracked in the classical Pauli frame and absorbed into the bases of subsequent measurements \cite{Omkar2022,Bartolucci2023}. 

Physically, they can also be implemented deterministically in dual-rail encoding: the Pauli-$X$ gate corresponds to a swap of the two modes encoding the qubit, while the Pauli-$Z$ gate corresponds to a phase shift by an angle $\pi$ on one of the two modes encoding the qubit.

\end{document}